\documentclass[twocolumn,showpacs,showkeys,preprintnumbers,pre,amsmath,amssymb,superscriptaddress]{revtex4}
\usepackage{graphicx}
\usepackage{natbib}
\usepackage{xcolor}
\usepackage{amsmath}
\usepackage{enumerate}
\usepackage{hyperref}
\usepackage[toc,page]{appendix}
\usepackage[normalem]{ulem}
\def\strutdepth{\dp\strutbox}
\def\nw#1{\strut\vadjust{\kern-\strutdepth\vtop to0pt{\vss\hbox to\hsize
{\hskip\hsize\hskip5pt$\leftarrow$\hss\strut}}}{\em #1}}

\begin{document}

\title{Metastability at the Yield-Stress Transition in Soft Glasses}

\author{Matteo Lulli}
\altaffiliation{Present address: Department of Applied Physics, Hong Kong Polytechnic University, Hong Kong, China.}
\email{matteo.lulli@roma2.infn.it, matteo.lulli@gmail.com}
\affiliation{Dipartimento di Fisica, Universit\`a di Roma  ``Tor Vergata'' and INFN, Via della Ricerca Scientifica, 1 - 00133  Roma, Italy\\}
\author{Roberto Benzi}
\email{benzi@roma2.infn.it}
\affiliation{Dipartimento di Fisica, Universit\`a di Roma  ``Tor Vergata'' and INFN, Via della Ricerca Scientifica, 1 - 00133  Roma, Italy\\}
\author{Mauro Sbragaglia}
\email{sbragaglia@roma2.infn.it}
\affiliation{Dipartimento di Fisica, Universit\`a di Roma  ``Tor Vergata'' and INFN, Via della Ricerca Scientifica, 1 - 00133  Roma, Italy\\}

\date{\today}

\begin{abstract}
We study the solid-to-liquid transition in a two-dimensional fully periodic soft-glassy model with an imposed spatially heterogeneous stress. The model we consider consists of droplets of a dispersed phase jammed together in a continuous phase. When the peak value of the stress gets close to the yield stress of the material, we find that the whole system intermittently tunnels to a metastable ``fluidized'' state which relaxes back to a metastable ``solid'' state by means of an elastic-wave dissipation. This macroscopic scenario is studied through the microscopic displacement field of the droplets, whose time-statistics displays a remarkable bimodality. Metastability is rooted in the existence, in a given stress range, of two distinct stable rheological branches as well as long-range correlations (e.g., large dynamic heterogeneity) developed in the system. Finally, we show that a similar behavior holds for a pressure-driven flow, thus suggesting possible experimental tests.
\end{abstract}

\maketitle
\section{Introduction}\label{sec:intro}
Soft amorphous materials\textemdash such as emulsions, microgels, foams and colloidal suspensions\textemdash display a solid-to-liquid transition for sufficiently large values of an external forcing: they are solid at rest and able to store energy via elastic deformations, whereas they flow whenever the stress is above a critical threshold known as the yield stress~\cite{Larson}. The complex spatio-temporal behavior shown by soft-glasses at the yield-stress transition has been the subject of intense scrutiny in the recent years~\cite{Coussot,MansardColin12,BalmforthREVIEW,BonnReviewYIELDSTRESS}. Some materials, often denoted as ``simple'' yield-stress fluids~\cite{BonnReviewYIELDSTRESS} (e.g. microgels~\cite{Geraud13}, nonadhesive emulsions~\cite{Goyon08,Goyon10}) exhibit yielding properties which are rather homogeneous in space: For any imposed shear rate, even a small one, there is always a stress at which these materials can fluidify homogeneously; the steady flow dynamics is also typically preceded by a nontrivial transient behavior~\cite{Divoux10,Divoux11}. In other materials with thixotropic properties~\cite{BonnReviewYIELDSTRESS}, like adhesive emulsions~\cite{Becu06}, a specific kind of heterogeneous flow can be steadily established: If an imposed shear rate is smaller than a given threshold, the system may decompose in two distinct spatial regions, showing a solid and fluidized behavior respectively. By changing the shear rate value, the widths of the two regions are changed, whereas the shear stress remains constant. This phenomenon is known as shear banding~\cite{Olmsted08,Ovarlez09,SchallREVIEW,Fielding2,Divouxetal16,Fielding16,ourSM16}. Here, the term ``shear banding'' as a form of heterogeneous flow characterized by shear localization \emph{independently} of any stress heterogeneity \cite{BonnReviewYIELDSTRESS}. This differs from the shear localization induced by stress heterogeneity, where part of the material is above yield and part below; it also differs from the shear localization emerging in presence of slippage at the walls.

From the theoretical point of view, different phenomenological models have been proposed to capture the fundamental physics underlying soft-glasses behaviors. In some cases [such as the soft-glassy-rheology (SGR) model~\cite{Sollich97,Sollich98,Sollich2000} or shear-transformation-zone (STZ) theory~\cite{Falk98}] the notion of ``effective temperature'' provides a useful way to describe the onset of the plastic flow in soft glasses. Such ``temperature'' is actually thought of as a quantification of the mechanical noise induced by the flow itself~\cite{Sollich97,Sollich98,Sollich2000} and triggers activated hopping through the energy landscape of the system. Moreover, it has been clearly demonstrated both experimentally~\cite{Chikkadi2011,Knowlton14,Linetal15} and numerically~\cite{Tigheetal17} that soft glasses exhibit a nontrivial size dependence. This may give rise to ``nonlocal'' rheological effects~\cite{Goyon08,Goyon10} parameterized by a cooperativity length~\cite{Goyon08,KEP09,Goyon10,Geraud13} estimating the typical size of the region involved in plastic rearrangements of the constituents following local elastic deformations. A recent proposal~\cite{ourSM16} has also linked cooperativity effects and nonlocal rheology to the emergence of shear-banding configurations. From a more general perspective, the shear-banding phenomenon has often been interpreted as the signature of a dynamic transition with a ``phase coexistence'' of two distinct states in space~\cite{Varnik03,Picard02,KEP09}: a jammed solid state and a fluidized state. A common explanation is to assume an underlying nonmonotonous rheological curve relating the stress to the shear rate~\cite{Olmsted08,Fielding2,Divouxetal16}, with two stable branches separated by an unstable branch. This nonmonotonicity has also been linked to the competition between different timescales related to different physical processes~\cite{Picard02,Coussot02,CoussotOvarlez10,Martens12} (e.g. aging vs. flow-induced rejuvenation in Ref.~\cite{Picard02} or restructuring time vs. stress-release time in Ref.~\cite{CoussotOvarlez10}). When the minimum of the rheological curve occurs at very small shear rates, one can draw a ``simple'' picture of coexisting branches~\cite{BonnReviewYIELDSTRESS}: a {\it solid} branch described by zero shear ($S=0$) and stress $\sigma$ in the interval $[0,\sigma_{\mbox{\tiny{st}}}]$, where $\sigma_{\mbox{\tiny{st}}}$ is referred to as the static yield stress; and a fluidized branch characterized by an Herschel-Bulkley (HB) relation of the type $\sigma = \sigma_{\mbox{\tiny{Y}}} + A S^n$~\cite{HB}, with $\sigma_{\mbox{\tiny{Y}}}<\sigma_{\mbox{\tiny{st}}}$ denoted as the dynamic yield stress. For stress values $\sigma \in [\sigma_{\mbox{\tiny{Y}}}, \sigma_{\mbox{\tiny{st}}}] $ the shear rate is multivalued, hence the phase coexistence in space. For shear rate $S$ greater than the critical shear $S_c= ((\sigma_{\mbox{\tiny{st}}}-\sigma_{\mbox{\tiny{Y}}})/A)^{1/n}$, the rheology of the system is described uniquely by the HB relation and no shear-banding is observed. This scenario has been explored and discussed in glassy models and numerical simulations~\cite{Berthier2003,Varnik03,Varnik04,XuOhern06,Chaudhuri12}.

In this paper we want to look at the statistical properties of the yield-stress transition when $\sigma_{\mbox{\tiny{Y}}}<\sigma_{\mbox{\tiny{st}}}$ from a different point of view. Permanent shear bands are often observed by applying an external velocity difference, say $\Delta U$ on a system of size $L$~\cite{ourSM16}. For $\Delta U/L < S_c$ the system shows a homogeneous stress in space and splits into two shearing regions (a solid and a fluidized band) which permanently persist in time. Now, let us consider the same system under an imposed space-dependent stress ranging, say, from $0$ to some value $\sigma_{\mbox{\tiny{p}}}$ close to $\sigma_{\mbox{\tiny{st}}}$. In this case, we have two solutions linked to the two possible branches. If the rate of plastic rearrangements is large enough, the system can perform activated processes and transitions \emph{in time} between the two solutions may be observed. In other words, for a relatively narrow range of values of the imposed shear stress peak $\sigma_{\mbox{\tiny{p}}}$, one should be able to observe a clear bimodality in the probability distribution of a global rheological variable, like the space-averaged velocity, or some other convenient observable. Hence, we expect a time bimodality because of the repeated (back-and-forth) transitions between two different states which are unimodal in space. Such transitions are expected to be enhanced by the choice of a heterogeneous stress field which reduces the extent of the spatial region in which transitions take place. Based on numerical simulations of a soft-glassy model~\cite{ourEPL10,ourSM12,ourEPL13,ourSM14,ourJFM15,ourEPL16} (see Sec.~\ref{sec:model}) we aim at providing a clear evidence that the above scenario holds.

In Sec.~\ref{sec:kolmogorov} we will analyze the rheological response at ``large scales'' and analyze the signatures of bimodality in the time evolution of the flow; then, in Sec.~\ref{sec:kolmogorovb} we enrich these observations with a comprehensive analysis of the rheological response at ``small scales'', i.e., by studying the statistical properties of the displacement field of the microstructural constituents. When bimodality is observed, we also observe that the overlap-overlap correlation length (see Sec.~\ref{sec:discussion}) becomes of the same order of the system size. We argue that a long-range correlation function among plastic events is necessary in order to observe transitions in time from one state to the other. Preliminary investigations for a pressure-driven flow (see Sec.~\ref{sec:confined}) will also support the same scenario, thus suggesting an experimental setup that could be used to test the predictions of numerical simulations. Some concluding remarks will be given in Sec.~\ref{sec:conclusions}. We believe that our results open a new perspective in the phenomenology of shear-banding in soft glasses.

\section{Model}\label{sec:model}
\begin{figure}[!t]
\includegraphics[scale=0.425]{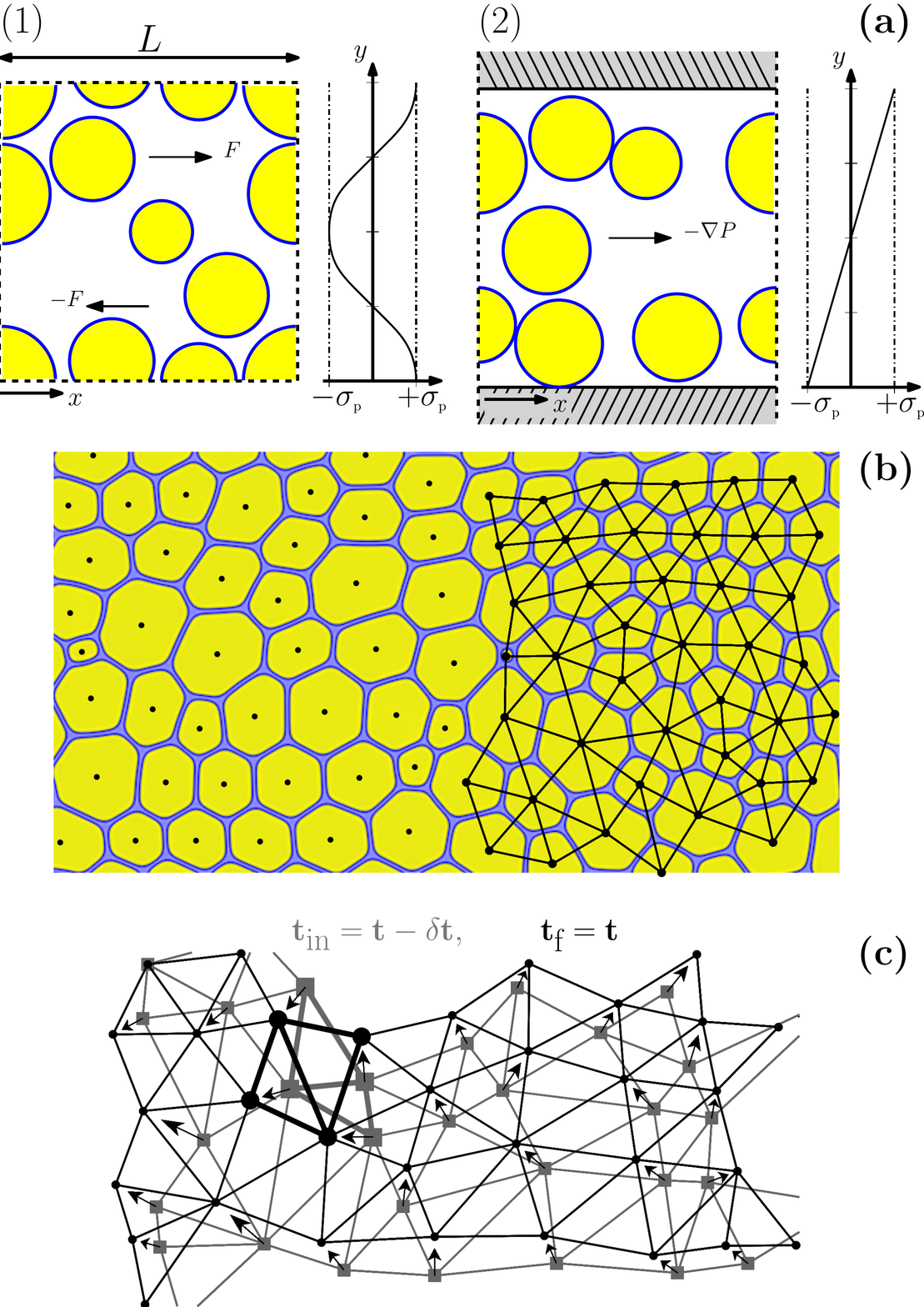}
\caption{\label{fig:sketch} Sketch depicting the fundamental quantities used in our analysis. (a) Flow setups and relative stress profiles: (1)  Kolmogorov flow on fully periodic square domain of size $L$; (2) Pressure-driven flow with stream-wise periodic boundary on a square domain. (b) Density field of the simulated soft-glassy model: deformable droplets in yellow are jammed together in a continuous phase (light blue); the droplets' centers of mass are indicated with a dot and are connected in their Delaunay triangulation \cite{Delaunay,ourGPU16} in the right half of the panel. (c) Comparison between two successive Delaunay triangulations at initial, $t_{\mbox{\tiny{in}}}$ (gray color with squared points), and final time $t_{\mbox{\tiny{f}}}$ (black color with round points): Arrows indicate the value of the displacement field $\vec{\Delta}_i(t)$ [see Eq.~\eqref{eq:Delta}] at each droplet; the region where a plastic rearrangement occurs (i.e., an edge-flip in the triangulation~\cite{ourGPU16}) is highlighted with thicker lines.
}
\end{figure}

We simulated a soft-glassy model by means of a lattice Boltzmann (LB) equation which allows the simulations of droplets of one component dispersed in another component~\cite{ourJCP09,ourEPL10,ourSM12,ourEPL13,ourSM14,ourJFM15,ourEPL16}. Droplets are stabilized against coalescence (see Fig.~\ref{fig:sketch}) by the combined effect of attractive and repulsive interactions~\cite{ourJCP09}. In previous publications we showed that the model displays many of the well-known properties observed for soft glasses. Importantly,  for shear-controlled experiments (i.e., in Couette geometry) it behaves as a non-Newtonian fluid displaying a dynamic yield stress $\sigma_{\mbox{\tiny{Y}}}$~\cite{ourEPL10}, a nonlinear HB rheology with HB exponent $n \simeq 0.5$~\cite{ourSM14}, elastic shear waves and plastic rearrangements~\cite{ourSM15}. For values of the stress $\sigma$ larger than $\sigma_{\mbox{\tiny{Y}}}$, our model shows quantitative agreement with nonlocal rheology theories~\cite{ourSM12,ourJFM15} which have been used to rationalize the flow of concentrated emulsions~\cite{Goyon08,Goyon10,KEP09} and other yield-stress fluids~\cite{Geraud13,katgert10,ourJFM15} in confined geometries. The values of the cooperativity scale $\xi$ extracted from the model \cite{Derzsi17,Derzsi18,ourSM14} are in agreement with experimental observations~\cite{Goyon08,Bonn15}. Recently, the model has been used in synergy with experiments on real emulsions in order to quantify the impact of the fluidization induced by the roughness of microchannels on the flow behavior of the emulsion \cite{Derzsi17,Derzsi18}. As shown in the reminder of the paper, for this ``model emulsion'' the static and dynamic yield-stress values are found to differ. Looking at the literature on real emulsions, we know that ``pure'' emulsions do not show this behavior, whereas loaded ``attractive'' emulsions actually do~\cite{Fall10,Molleretal09,Chaudhuri12}. Hence, in terms of the yielding properties, our model bears similarities with the behavior of an ``attractive'' emulsion with $\sigma_{\mbox{\tiny{Y}}}<\sigma_{\mbox{\tiny{st}}}$. Hereafter, we present all our numerical results by rescaling the LB units in such a way that the flowing rheological branch for a Couette forcing is given by $\sigma/\sigma_Y = 1+ S^{1/2}$, where $S$ is the shear. The system we consider is two dimensional, with $x$ and $y$ being the streamwise and spanwise coordinates respectively. We will study the rheology of our model by imposing a space-dependent stress.
For this purpose, we consider fully periodic boundary conditions with a space-dependent forcing imposing the $xy$ component of the stress (Kolmogorov flow):
\begin{equation}
\sigma_{xy}(x,y) = \sigma_{\mbox{\tiny{p}}} \cos\left (\frac{2\pi}{L}y\right),
\end{equation}
where $L$ is the system size which has the same value in both directions and $\sigma_{\mbox{\tiny{p}}}$ is the peak value for the stress (see Fig.~\ref{fig:sketch}). A very similar setting has been used in previous experimental~\cite{Tigheetal17} and numerical~\cite{KawasakiBerthier2016,Chaudhuri_prl12} works. The choice of a fully periodic setup is initially taken in order to avoid possible wall effects and dependence on boundary conditions, which may alter the rheological response of the system~\cite{Barnes,Buscall,Gibaud08,Seth12,Bonn15}.
Later, in section~\ref{sec:confined} we will discuss some preliminary simulations for a pressure-driven flow. In the fully periodic setup, for a Newtonian fluid with constant viscosity $\eta$, the streamwise component of the stationary velocity field induced by the stress would read
\begin{equation}\label{eq:ux}
u^{\mbox{\tiny{N}}}_x(x,y) = u_0^{\mbox{\tiny{N}}}\sin\left(\frac{2\pi}{L}y\right)
\end{equation}
where the peak value for the Newtonian velocity profile $u_0^{\mbox{\tiny{N}}}=\frac{L}{2 \pi \eta}\sigma_{\mbox{\tiny{p}}}$ is a constant. In the model, an important control parameter is the quantity $R=2\delta \sqrt{N}/L$ where $\delta$ is the average thickness of the continuous phase, $N$ is the number of droplets and $L$ is the system size. Such a quantity is a measure of the ratio between the interface area and the area occupied by the droplets. Note that $1-R$ should be considered proportional to the packing fraction in our system. The numerical simulations for the Kolmogorov flow have been performed with $L=1024$ grid points, $N=512$ droplets and $R = 0.09$ which implies a packing fraction well above the jamming point.
\section{Rheological response at ``large scales''}\label{sec:kolmogorov}
The simplest way to measure the rheology in our system is to compute the characteristic shear $S$ as a function of $\sigma_{\mbox{\tiny{p}}}$. The value of $S$ is computed using the average streamwise velocity profile $u_x(y,t) = L^{-1} \sum_x u_x(x,y,t) $ at time $t$ and performing its projection onto the viscous profile in Eq.~\eqref{eq:ux}\begin{equation}
u_s(t) = \frac{2}{L}\sum_{y=0}^{L-1} u_x(y,t)\sin\left(\frac{2\pi}{L}y\right).\label{eq:viscousProj}
\end{equation}\begin{figure}[t]
\includegraphics[scale=0.76]{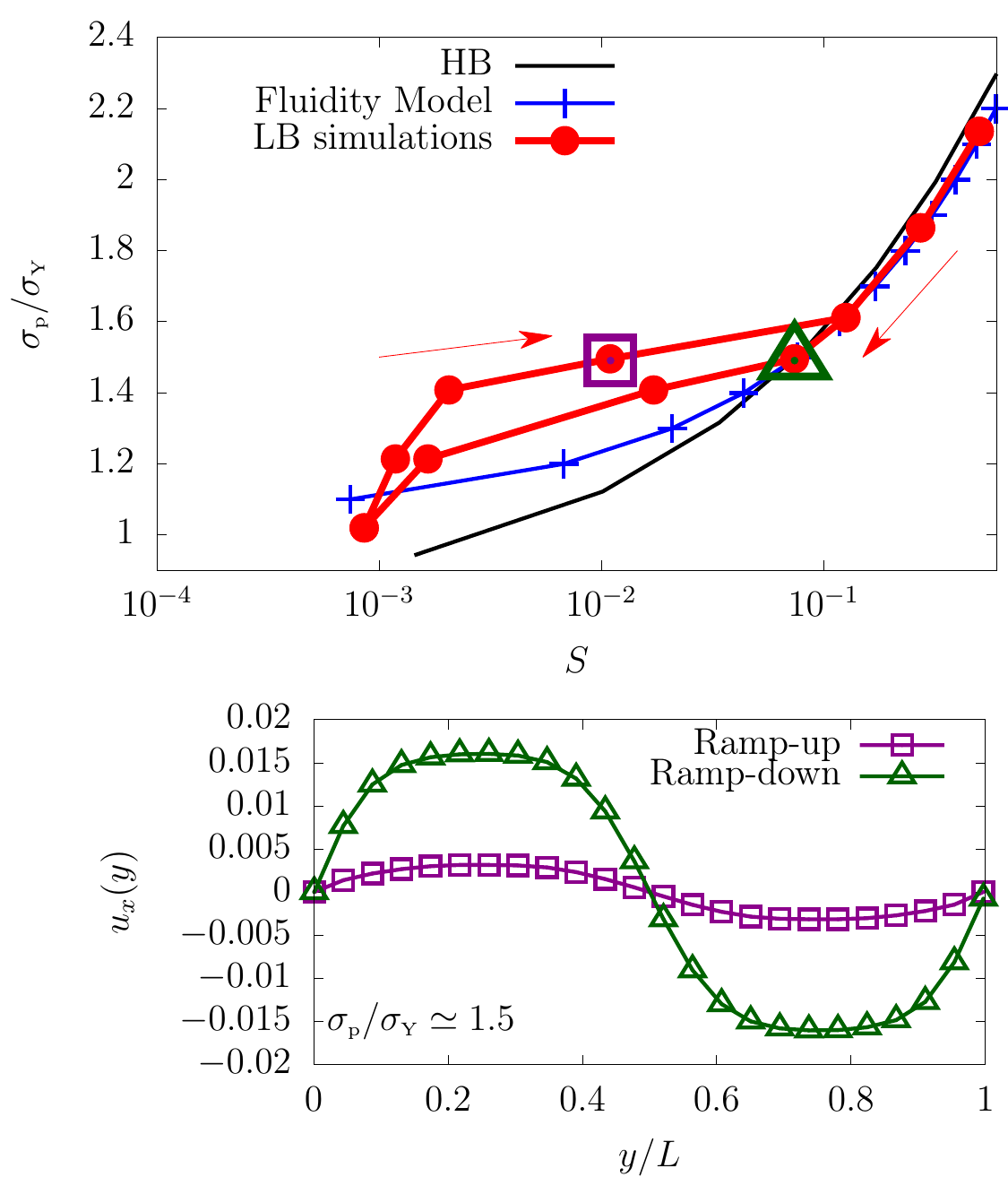}
\caption{\label{fig:rheology} Top panel: Rheology data for the Kolmogorov flow setup. The hysteresis cycle for the LB simulations (red dots) is clearly visible: ramp-up (upper line with right pointing arrow) and ramp-down simulations (lower line with left pointing arrow). Rheological data extracted from simulations are compared to the results obtained from the fluidity model~\cite{Goyon08,KEP09,Goyon10} (blue crosses) and the Herschel-Bulkley (HB)~\cite{HB} fit in a Couette geometry (black continuous line). Neither models can describe the solid branch and the transition to the plastic one; however, the fluidity model better describes the flowing regime. The purple empty square and the green empty triangle signal the position on the rheological curve of the corresponding profiles shown in the bottom panel.  Bottom panel: Velocity profiles for the Kolmogorov flow at fixed peak-stress value $\sigma_{\mbox{\tiny{p}}}/\sigma_{\mbox{\tiny{Y}}}\simeq 1.5$ obtained from two different protocols: ramping up from lower peak-stress values (purple empty squares) and ramping down from larger peak-stress values (green empty triangles). This is clear evidence of the existence of two stable rheological branches signaling that the static yielding threshold is above the dynamic one~\cite{Varnik03}.}
\end{figure}From $u_s(t)$ we compute $s(t)=2\pi u_s(t)/L$, whose time average provides the value of the shear $S$. 
In the top panel of Fig.~\ref{fig:rheology} we show the rheological curve obtained in our system: Starting from $\sigma_{\mbox{\tiny{p}}}/\sigma_{\mbox{\tiny{Y}}} \simeq 1$ we perform a series of numerical simulations (red bullets) by increasing stepwise the peak stress $\sigma_{\mbox{\tiny{p}}}$ (i.e. ``ramp-up'' protocol). At relatively large values of the forcing (namely for $\sigma_p/\sigma_Y \simeq 2.2$) the system is completely fluidized. Next we reduce the forcing (i.e. ``ramp-down'' protocol) using exactly the same values $\sigma_{\mbox{\tiny{p}}}/\sigma_{\mbox{\tiny{Y}}}$ of the ``ramping-up'' simulations: As observed in the top panel of Fig.~\ref{fig:rheology} a clear (although small) hysteresis loop is observed. In the top panel of Fig.~\ref{fig:rheology} the black continuous line refers to the same quantities for a simple HB fluid whose parameters are the same as those observed for our model in a Couette geometry~\cite{ourSM14} while the blue connected crosses refer to the same HB fluid supplemented with cooperativity effects, obtained using the {\it steady} nonlocal fluidity model~\cite{Goyon08,KEP09,Goyon10}. Finally, in the lower panel of Fig.~\ref{fig:rheology} we show the average velocity profiles $u_x(y)$ observed at $\sigma_{\mbox{\tiny{p}}}/\sigma_{\mbox{\tiny{Y}}} \simeq 1.5$ for the ``ramp-up'' simulation (purple squares) and ``ramp-down'' simulation (green triangles). Velocity profiles $u_x(y)$ are obtained from an average in time of $u_x(y,t)$. The results shown in Fig.~\ref{fig:rheology} clearly demonstrate the existence in our system of two rheological branches with a dynamical yield stress smaller than the static one. Moreover, looking at the top panel of Fig.~\ref{fig:rheology} we can immediately observe that the yielding point is above the yielding threshold evaluated in homogeneous conditions, i.e. $\sigma_{\mbox{\tiny{p}}}/\sigma_{\mbox{\tiny{Y}}}\simeq 1.4$. Qualitatively, we can argue that this is a consequence of the nonlocality in the flow coupled to the heterogeneity of the stress. Indeed, for the flow to occur, the peak stress needs to be above $\sigma_{\mbox{\tiny{Y}}}$ in a spatial region of the order of the cooperativity length~\cite{Chaudhuri_prl12,ourEPL13}. The net effect of this is to increase the yielding threshold. However, a closer quantitative inspection reveals that the nonlocal model works very well only when the peak stress is well above the yield stress, while it fails to describe the transition point for $\sigma_{\mbox{\tiny{p}}}/\sigma_{\mbox{\tiny{Y}}}\simeq 1.4$. We indeed observe an abrupt transition in the rheological response that neither the simple HB model nor the stationary nonlocal fluidity model are able to capture. This contrasts with previous observations in yield-stress fluids subject to heterogeneous stress distribution~\cite{Chaudhuri_prl12}. We are therefore interested in investigating the nature and properties of this transition.

\begin{figure}[!b]
\includegraphics[scale=0.81]{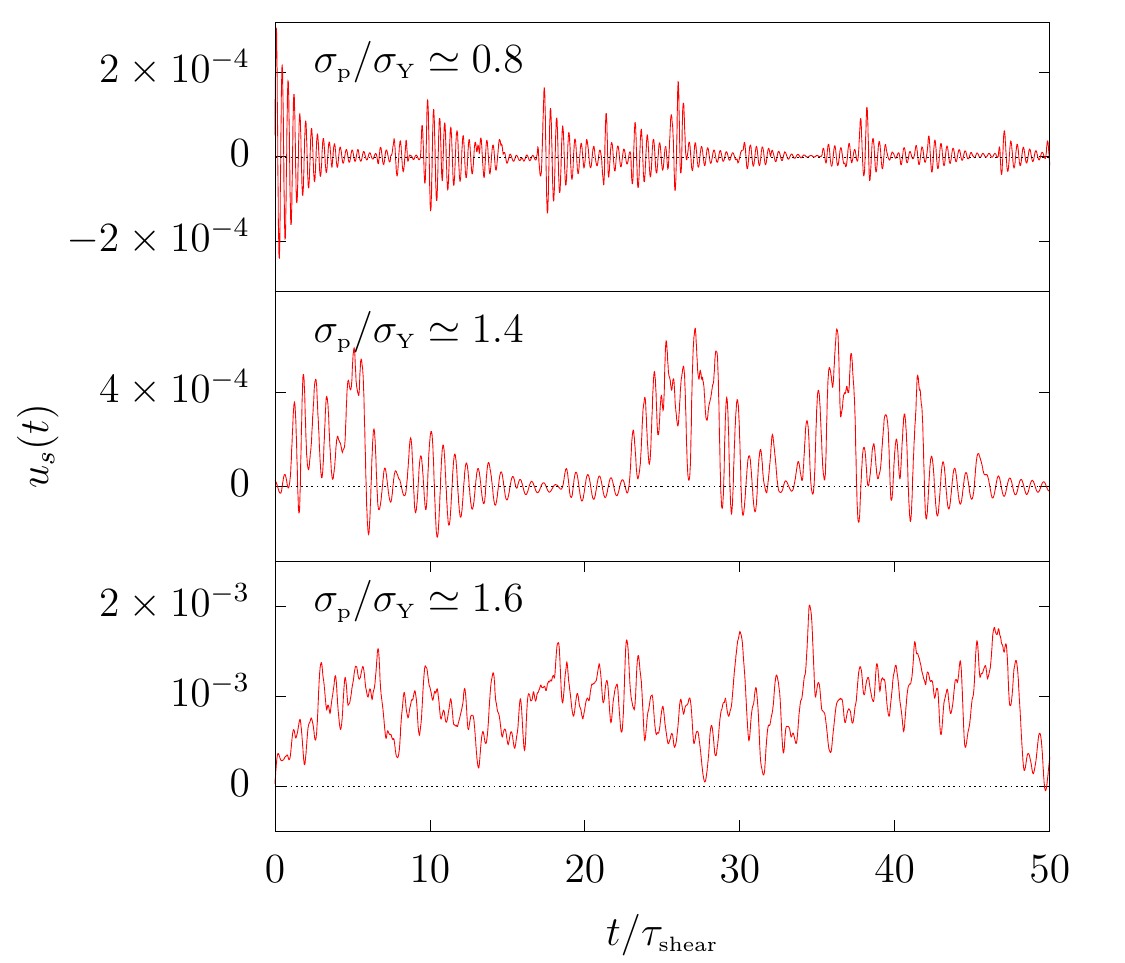}
\caption{\label{fig:three_Us} Time sequence for the projection of the velocity field onto the viscous solution $u_s(t)$ [see Eq.~\eqref{eq:viscousProj}] for Kolmogorov flow simulations at different peak stress $\sigma_{\mbox{\tiny{p}}}$ values. Time is rescaled by the stress-dependent shear time $\tau_{\mbox{\tiny{shear}}} = S^{-1}$ as derived from Eq.~\eqref{eq:viscousProj}. For small forcing, $\sigma_{\mbox{\tiny{p}}}/\sigma_{\mbox{\tiny{Y}}} \simeq 0.8$, the system responds elastically and dissipates mainly through elastic waves visible from the periodic oscillations around $0$. At $\sigma_{\mbox{\tiny{p}}}/\sigma_{\mbox{\tiny{Y}}} \simeq 1.4$, the systems is at the middle point between the two branches (see Fig.~\ref{fig:rheology}) and it intermittently switches between an elastic response and a plastic flowing regime for which $u_s(t) > 0$. At $\sigma_{\mbox{\tiny{p}}}/\sigma_{\mbox{\tiny{Y}}} \simeq 1.6$ the system is plastically flowing~\cite{Goyon08,KEP09,Goyon10}.  
}
\end{figure}

To get an intuitive picture on the system behavior at the transition, we show in Fig.~\ref{fig:three_Us} the time behavior of $u_s(t)$ for three different values of $\sigma_{\mbox{\tiny{p}}}$. All of the following simulations have been performed using the ramp-up protocol unless explicitly stated otherwise (see Section~\ref{sec:confined}). For relatively small $\sigma_{\mbox{\tiny{p}}}$ (top panel) the system intermittently tries to flow with an average value of $u_s$ close to zero; at large $\sigma_{\mbox{\tiny{p}}}$ (lower panel) the system is fluidized, and the signal corresponds to a plastic flow, as expected. The interesting point is the behavior of the system at $\sigma_{\mbox{\tiny{p}}}/\sigma_{\mbox{\tiny{Y}}}\simeq 1.4$ (middle panel): The system persists for relatively long time in a fluidized state and then goes back in a ``solid'' state.
We also notice in the upper and middle panels of Fig.~\ref{fig:three_Us} strong periodic oscillations of $u_s(t)$. These oscillations are due to elastic waves generated in the system. The signal shown in the upper panel recalls the ``stick-slip'' behavior observed near the yield-stress transition in shear controlled systems~\cite{Varnik03}:  since we impose the stress, the shear (or the velocity) shows intermittent bursts of activity. It is much less immediate, however, to understand the physics behind the behavior of $u_s(t)$ shown at $\sigma_{\mbox{\tiny{p}}}/\sigma_{\mbox{\tiny{Y}}}\simeq 1.4$. Since the intermittency in $u_s(t)$ is due to plastic rearrangements occurring in the system, it is important to inspect the system behavior at the scales of the microstructural constituents in order to get a deeper insight about the nature of the observed transition.
\section{Rheological response at ``small scales''}\label{sec:kolmogorovb}
Plastic rearrangements are localized topological changes in the droplets configurations. In our system, we can identify plastic rearrangements, corresponding to topological changes in the Voronoi tessellation of the centers of mass, by using its dual Delaunay triangulation (see Fig.~\ref{fig:sketch}): A plastic event happens whenever a link in the triangulation flips~\cite{ourGPU16}. Next, we need to measure the droplet displacement during plastic rearrangements and try to understand whether this measure can be correlated to the observations discussed in Fig.~\ref{fig:three_Us}. For this purpose, we start by looking at the displacement $\vec{\Delta}_i (t)$ of the droplets defined as 
\begin{equation}
\vec{\Delta}_i (t)=\vec{x}_i(t) - \vec{x}_i(t - \delta t),\label{eq:Delta}
\end{equation}
where $\vec{x}_i(t)$ is the position of the center of mass of the $i$-th droplet at time $t$ and $\delta t$ is a given time interval which in our simulations is set to be $\delta t=100$ simulation time steps. This choice corresponds roughly to $\delta t=t_{\mbox{\tiny drop}}/10$, where $t_{\mbox{\tiny drop}}=\eta\, \langle R \rangle /\gamma$ is the droplet time, with $\langle R \rangle$ the average radius and $\gamma$ the surface tension.

\begin{figure}[t]
\includegraphics[scale=0.79]{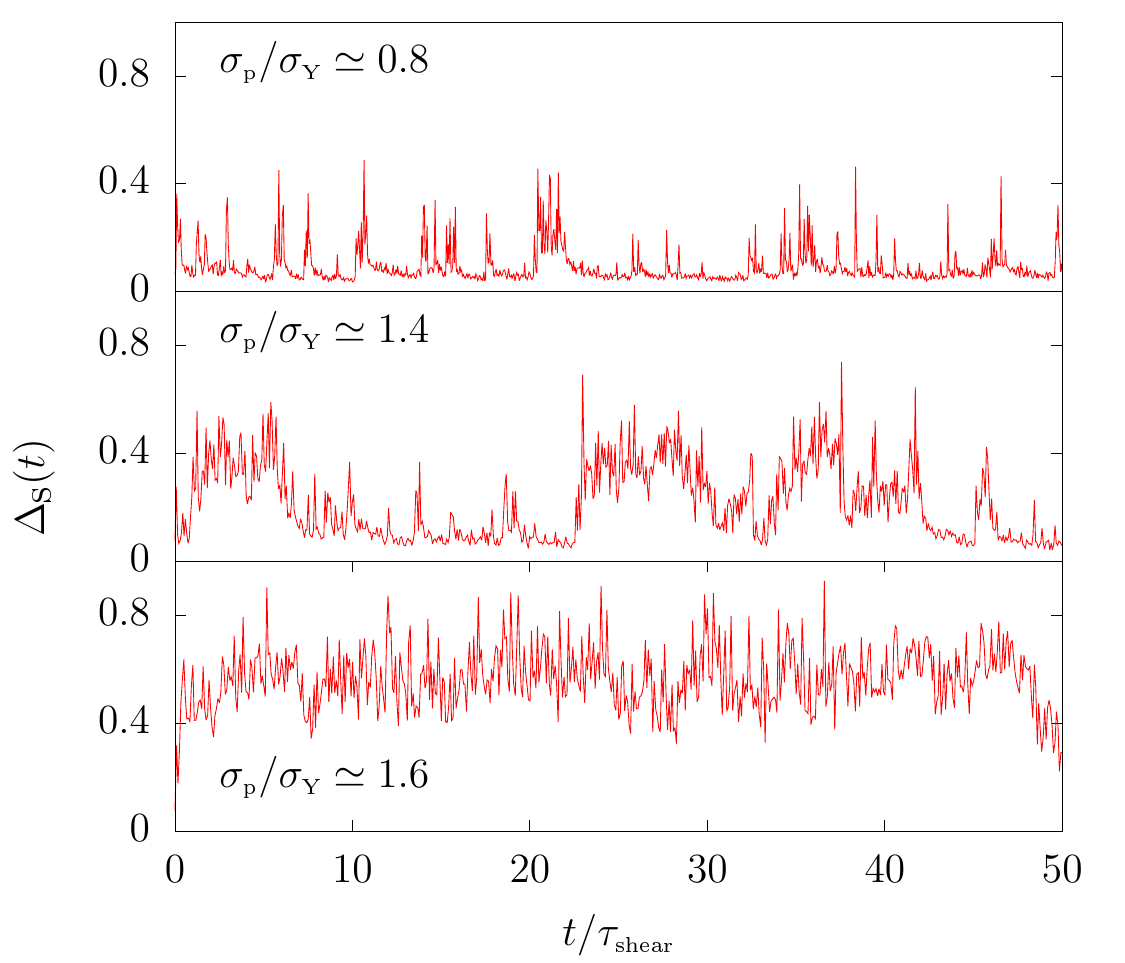}
\caption{\label{fig:three_Delta} Time sequence for the supremum of droplets displacements $\Delta_{\mbox{s}}(t)$ [see Eq.~\ref{eq:supremum}] at the same peak stresses displayed in Fig.~\ref{fig:three_Us} for the Kolmogorov flow. Time is rescaled by the stress-dependent shear time $\tau_{\mbox{\tiny{shear}}} = S^{-1}$ as derived from Eq.~\eqref{eq:viscousProj}. At the smallest forcing $\sigma_{\mbox{\tiny{p}}}/\sigma_{\mbox{\tiny{Y}}} \simeq 0.8$, $\Delta_{\mbox{s}}$ shows only a few intermittent spikes with the smallest absolute values. For a peak stress $\sigma_{\mbox{\tiny{p}}}/\sigma_{\mbox{\tiny{Y}}} \simeq 1.4$, passing from one rheological branch to the other (see Fig.~\ref{fig:rheology}), $\Delta_{\mbox{s}}$ displays both small and large stable values, whereas at $\sigma_{\mbox{\tiny{p}}}/\sigma_{\mbox{\tiny{Y}}} \simeq 1.6$ there are fluctuations around a large mean value. 
}
\end{figure}

As expected, $| \vec \Delta_i (t)|$ is a highly intermittent quantity both in $i$ (space) and time: It fluctuates around a small value when there are no plastic rearrangements, while it becomes large and strongly localized in space when a plastic rearrangement occurs somewhere in the system. For this reason, we consider 
\begin{equation}\label{eq:supremum}
\Delta_{\mbox{s}} (t) \equiv \sup_i |\vec{\Delta}_i(t)|,
\end{equation}
as a quantitative measure of plastic activity in the system. The behavior in time of $\Delta_{\mbox{s}}(t)$ is shown in Fig.~\ref{fig:three_Delta} for the same values of the peak stress discussed in Fig.~\ref{fig:three_Us}. Quite remarkably (but not surprisingly), the behavior of $\Delta_{\mbox{s}}(t)$ is qualitatively similar to the one shown by $u_s(t)$. However, an important difference must be stressed: $\Delta_{\mbox{s}}$ is not affected by the presence of elastic waves. This difference can be understood in a simple way: the displacement due to elastic waves is relatively small and it is coherent in space (all droplets oscillate); in contrast the displacement due to plastic rearrangements is rather large and not coherent in space. Therefore, our quantity $\Delta_{\mbox{s}}$ is not sensitive to elastic waves.
\begin{figure}[!hb]
\includegraphics[scale=0.76]{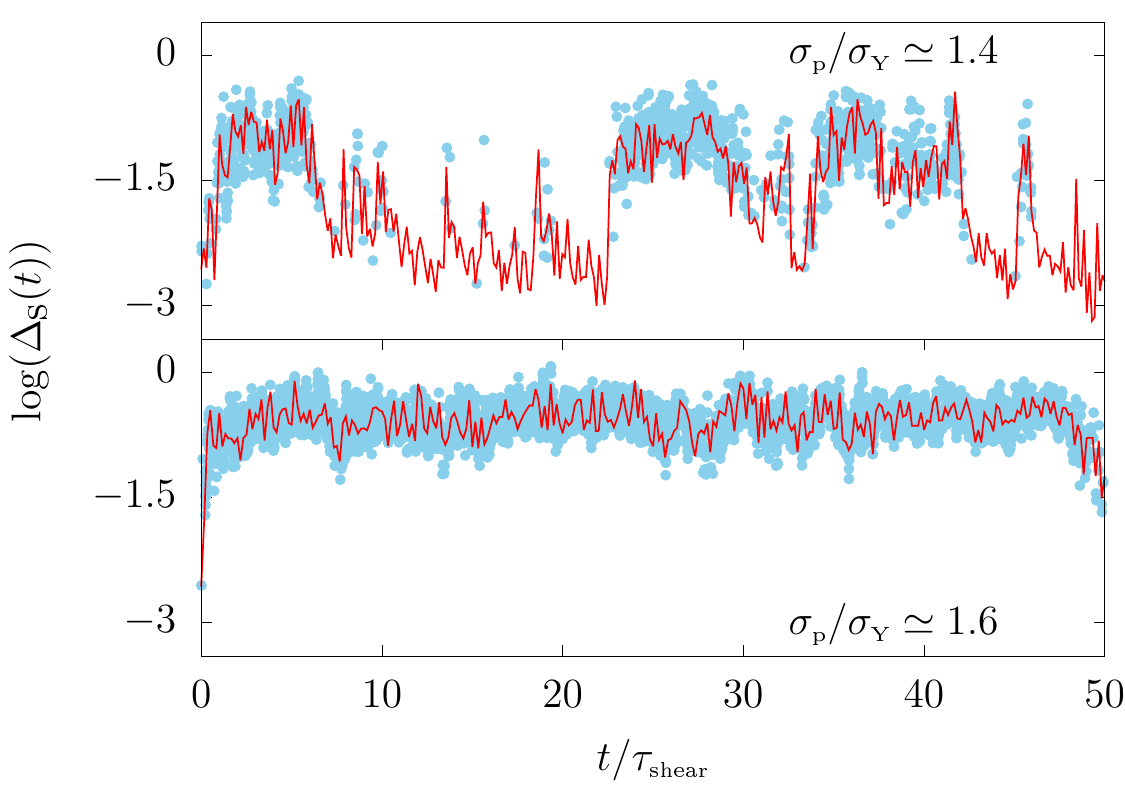}  
\caption{\label{fig:deltaRegimes} Lines represent data for $\log(\Delta_{\mbox{s}})$ while dots single out values concurrent with plastic rearrangements. Time is rescaled by the stress-dependent shear time $\tau_{\mbox{\tiny{shear}}} = S^{-1}$ as derived from Eq.~\eqref{eq:viscousProj}. Two different regimes are displayed. Top panel: $\sigma_{\mbox{\tiny{p}}}/\sigma_{\mbox{\tiny{Y}}}\simeq 1.4$ the system spends roughly half of the time in an elastic solid state and the other half in a plastic fluidized state where the plastic rearrangements cluster. Bottom panel: $\sigma_{\mbox{\tiny{p}}}/\sigma_{\mbox{\tiny{Y}}} \simeq 1.6$ the system is in a fluidized state and plastic rearrangements are homogeneously distributed.
}
\end{figure}

Finally, in Fig.~\ref{fig:deltaRegimes} we compare the amplitude of $\Delta_{\mbox{s}}$ and simultaneously track the time (blue dots) when plastic rearrangements occur. We show the time behavior of $\Delta_{\mbox{s}}$ for two different values of the peak stress: $\sigma_{\mbox{\tiny{p}}}/\sigma_{\mbox{\tiny{Y}}} \simeq 1.4$ showing the previously described intermittent behavior and $\sigma_{\mbox{\tiny{p}}}/\sigma_{\mbox{\tiny{Y}}} \simeq 1.6$ for which the system is plastically flowing~\cite{Goyon08,KEP09,Goyon10}. Inspection of Fig.~\ref{fig:deltaRegimes} suggests that we should consider the probability distribution $P(\log(\Delta_{\mbox{s}}))$, in agreement with the approach to intermittent fluctuations in dynamical systems theory \cite{BenziIntermittency1985}. We remark that, upon writing $Z = \log(\Delta_{\mbox{s}})$, it is easily shown that $P(Z) = \Delta_{\mbox{s}} P(\Delta_{\mbox{s}})$, i.e. the peak in the probability distribution of $P(\log(\Delta_{\mbox{s}}))$ corresponds to the relevant value of $\Delta_{\mbox{s}}$ contributing to the average $\langle \Delta_{\mbox{s}} \rangle$~\cite{TanguyLanforteBarrat}. 

\begin{figure}[!b]
\includegraphics[scale=0.79]{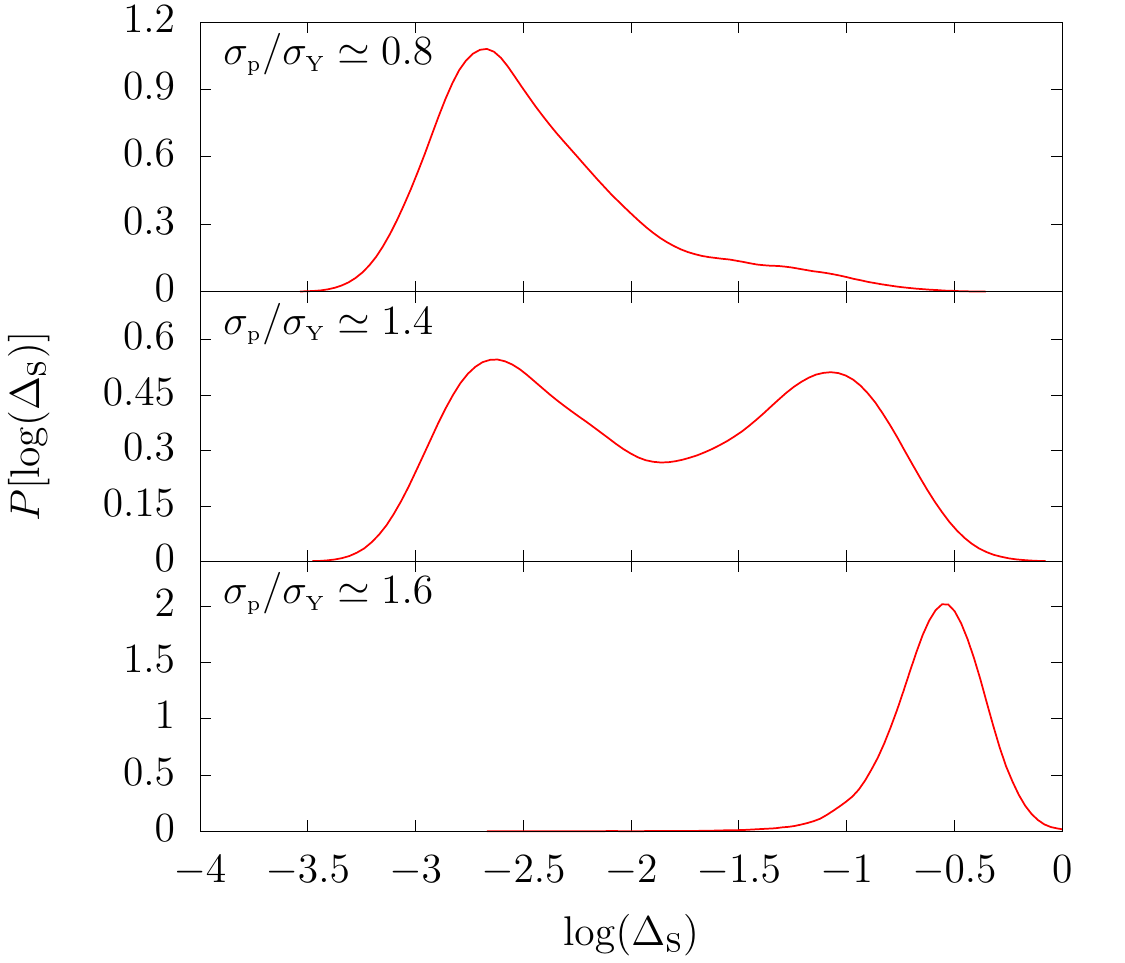}  
\caption{\label{fig:bim}Probability distribution functions for $\log(\Delta_{\mbox{s}})$ for three different values of the forcing displayed in Fig.~\ref{fig:three_Delta}. Top panel: At the smallest forcing, $\sigma_{\mbox{\tiny{p}}}/\sigma_{\mbox{\tiny{Y}}} \simeq 0.8$ there is one peak at small values with a long tail over larger values indicating the intermittent spikes of plastic activity the system experiences. Middle panel: at $\sigma_{\mbox{\tiny{p}}}/\sigma_{\mbox{\tiny{Y}}} \simeq 1.4$ the system spends time both in the solid elastic response branch (smaller peak) and in the plastic flowing one (larger peak; see Fig.~\ref{fig:rheology}) so that the probability distribution is bimodal. Bottom panel: at $\sigma_{\mbox{\tiny{p}}}/\sigma_{\mbox{\tiny{Y}}} \simeq 1.6$ only the fluidized state exists signaled by the peak at large values.}
\end{figure}

The probability distributions  $P(\log(\Delta_{\mbox{s}}))$ are shown in Fig.~\ref{fig:bim} for the three different peak stresses already considered before: at small $\sigma_{\mbox{\tiny{p}}}$, $P(\log(\Delta_{\mbox{s}}))$ is peaked at small values and shows a rather long tail; at large $\sigma_{\mbox{\tiny{p}}}$, $P(\log(\Delta_{\mbox{s}}))$  is peaked at large values corresponding to the plastic flow previously discussed. Remarkably, at the transition point $\sigma_{\mbox{\tiny{p}}}/\sigma_{\mbox{\tiny{Y}}} \simeq 1.4$, the probability distribution $P(\log(\Delta_{\mbox{s}}))$ is bimodal, i.e. the system shows transitions in {\it time} between two states with small (solid) and large (fluidized) values. Hence, we observe bimodality in time of two states that are unimodal in space.

Now, we go back to the results shown in Fig.~\ref{fig:three_Us}. The results discussed in terms of $\Delta_{\mbox{s}}$ 
\begin{figure}[t]
\includegraphics[scale=0.79]{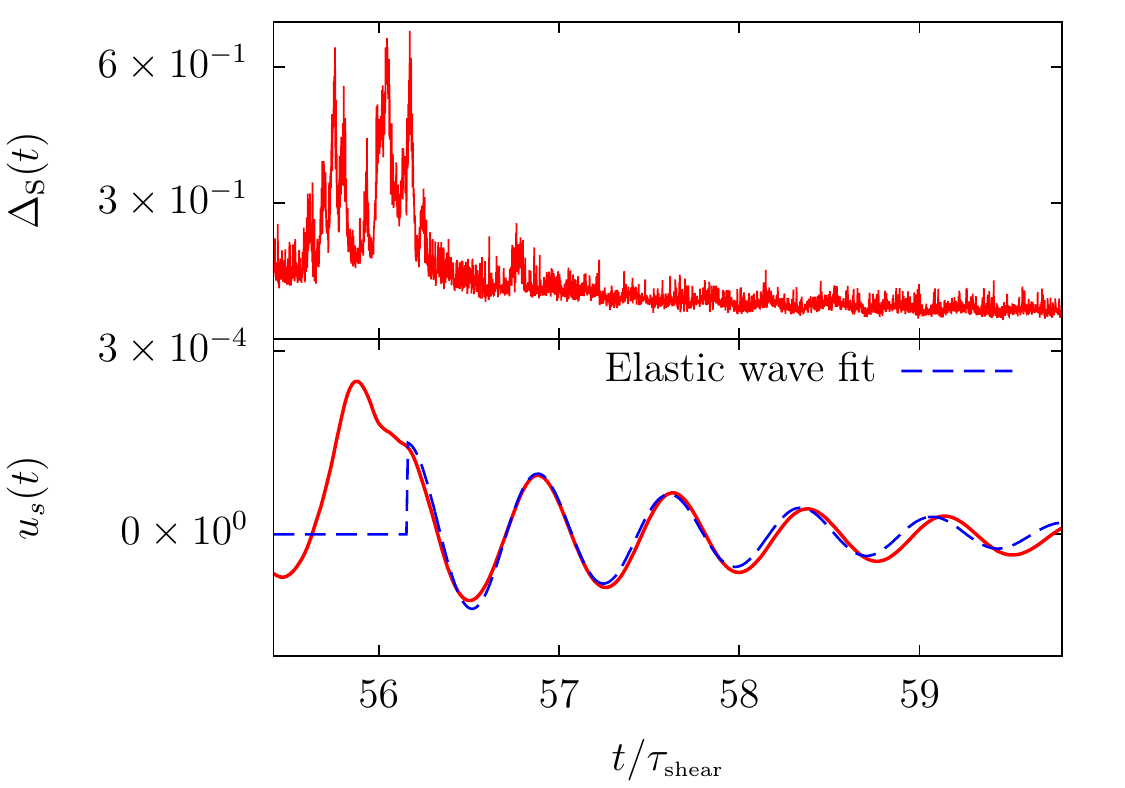}  
\caption{\label{fig:usVsD} Top panel: Time evolution of the quantity $\Delta_{\mbox{s}}(t)$ showing both plastic and elastic regimes. Bottom panel: Velocity field projection onto the viscous solution $u_s(t)$. It is possible to notice that both signals are rather compatible in the plastic regimes (high variability), whereas the elastic wave dissipation is clearly visible for $u_s(t)$ and practically does not affect the data for $\Delta_{\mbox{s}}(t)$.
}
\end{figure}
\begin{figure}[!hb]
\includegraphics[scale=0.77]{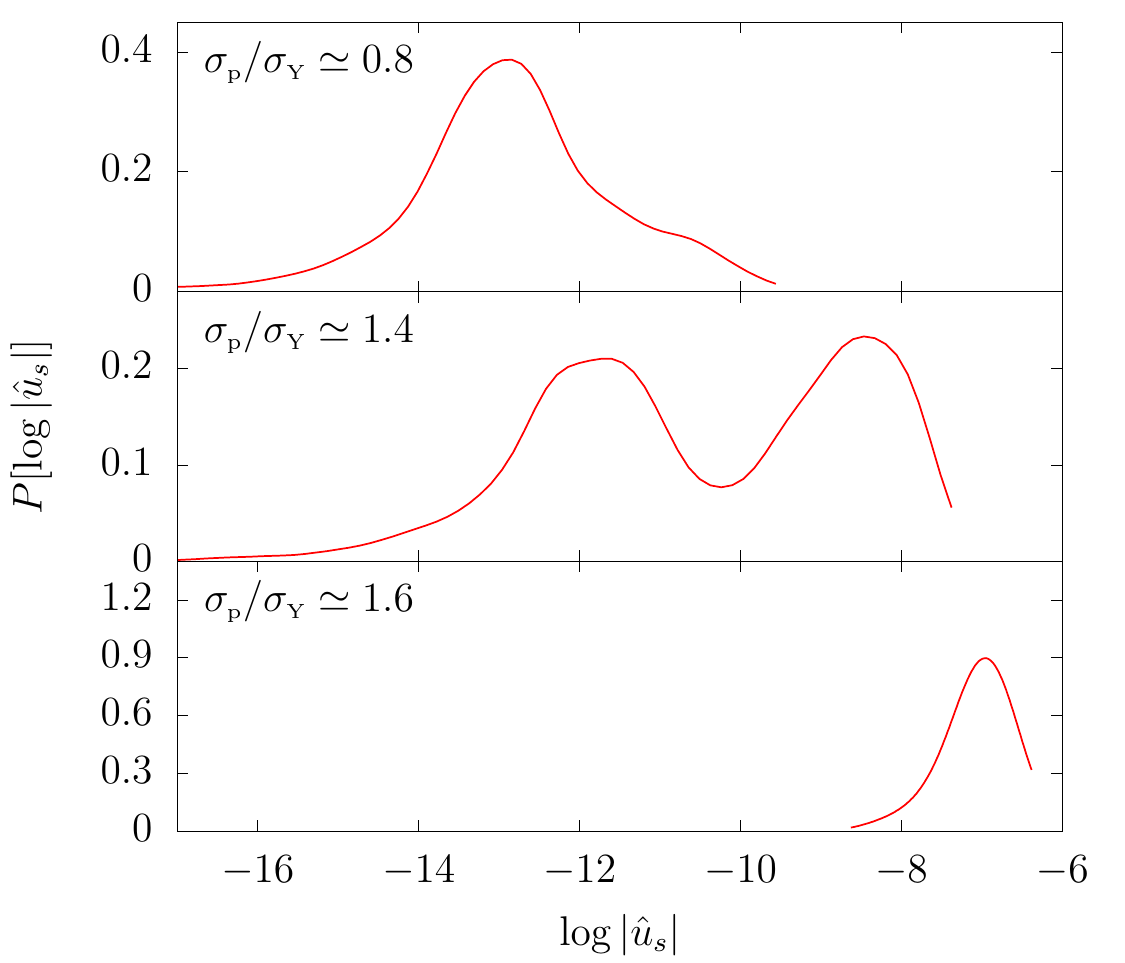}  
\caption{\label{fig:usBim} Probability distribution functions for $\log|\hat{u}_s(t)|$ for the same values of the peak stress $\sigma_{\mbox{\tiny{p}}}$ reported in Fig.~\ref{fig:bim}, where $\hat{u}_s(t)$ is the projection of the velocity field onto the viscous solution [see Eq.~\eqref{eq:viscousProj}] once elastic waves are filtered out [see Eq.~\eqref{eq:filter}]. We obtain the same qualitative behavior as in Fig.~\ref{fig:bim} stressing the transition from the solid branch (top panel) to the plastically flowing one (bottom panel) passing through a regime where both coexist (middle panel). See Fig~\ref{fig:rheology}.}
\end{figure}
suggest that transitions from the solid branch to the fluidized branch should be observed for $u_s$ as well. As already remarked, however, $u_s(t)$ is strongly perturbed by elastic waves which makes it impossible to observe the same bimodality unless the effects of elastic waves are removed. This can actually be done. In Fig.~\ref{fig:usVsD} we show a short snapshot of the time behavior of $\Delta_{\mbox{s}}$ (upper panel) and $u_s(t)$ (lower panel) for $\sigma_{\mbox{\tiny{p}}}/\sigma_{\mbox{\tiny{Y}}} \simeq 1.2$. When $\Delta_{\mbox{s}}$ becomes small, $u_s(t)$ shows damped oscillations near $u_s=0$. Knowing the period and the dissipation time of the elastic wave~\cite{ourSM14}, it is possible to fit the damped oscillations rather well as shown from the blue dashed line in the lower panel. We then obtain the filtered signal $\hat{u}_s(t)$ by computing a running average
\begin{equation}
\hat{u}_s(t) = \frac{1}{2T_{\mbox{\tiny{el}}}} \sum_{i = t - T_{\mbox{\tiny{el}}}}^{t+T_{\mbox{\tiny{el}}} - 1} u_s(i),\label{eq:filter}
\end{equation}
where $2T_{\mbox{\tiny{el}}}$ is the oscillation period of the elastic waves. In Fig.~\ref{fig:usBim} we report the probability distribution for $\log|\hat{u}_s|$ for different peak stresses. We consider the $\log|\hat{u}_s|$ for the same reasons previously discussed for $\Delta_{\mbox{s}}$. Comparing Figs.~\ref{fig:bim} and~\ref{fig:usBim}, once the elastic waves are filtered from the original signals, the probability distributions of $\log|\hat{u}_s|$ display the same features as $P(\log(\Delta_{\mbox{s}}))$ at the same forcing.

\section{Discussion}\label{sec:discussion}

\begin{figure}[!b]
\includegraphics[scale=0.77]{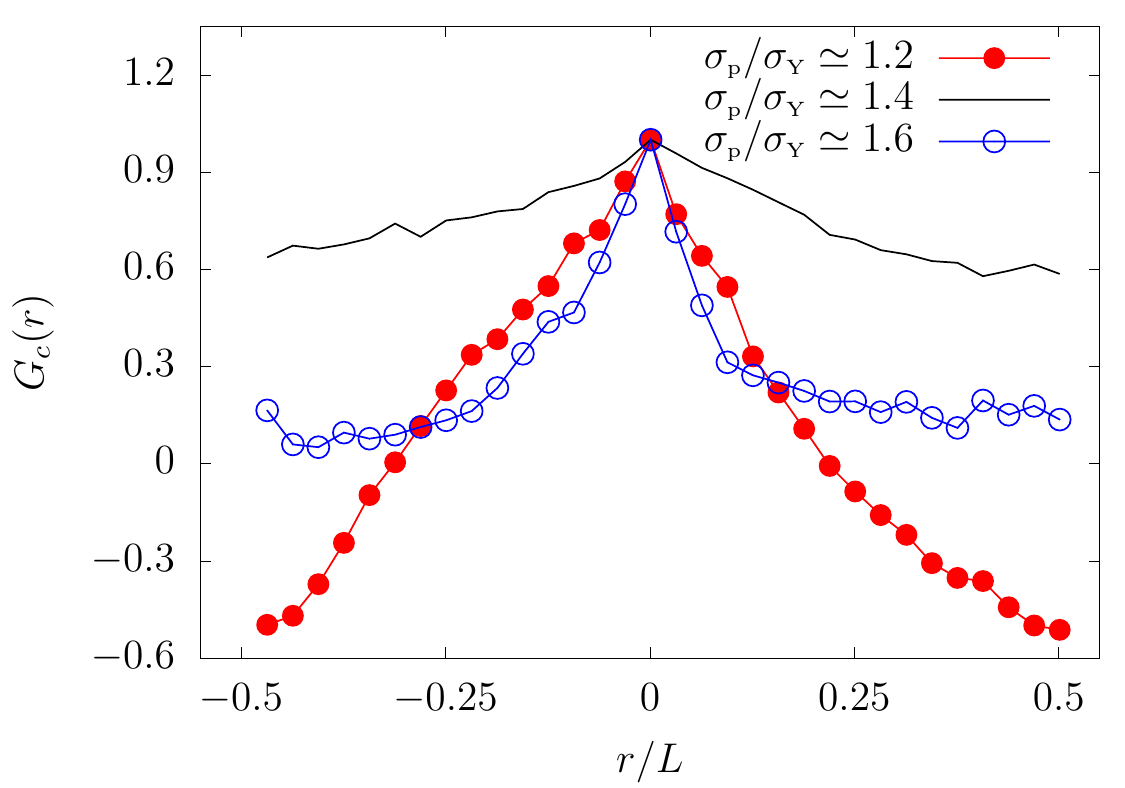}  
\caption{\label{fig:corr} Overlap-overlap connected correlation function $G(r)$ [See Eq.~\eqref{corrover} and text for details] for different values of peak stress $\sigma_{\mbox{\tiny{p}}}$. It is possible to notice that the correlation function takes on the highest values on the entire domain for the ratio $\sigma_{\mbox{\tiny{p}}}/\sigma_{\mbox{\tiny{Y}}} \simeq 1.4$ corresponding to the bimodal behavior (see Fig.~\ref{fig:bim}). The integral of $G(r)$ is usually known as dynamic heterogeneity~\cite{parisiFranz2000}.}
\end{figure}

Bimodal distributions and/or metastability have already been reported in the literature of amorphous systems~\cite{LamaitreCaroli,TanguyLanforteBarrat,Heussinger10,Chikkadi2014,KawasakiBerthier2016,Jaiswal2016,Procaccia17}. Regarding bimodal distributions, a recent theoretical work on amorphous solids by Jaiswal {\it et al.}~\cite{Jaiswal2016} showed bimodality for an order parameter \emph{ad hoc} constructed to see how much the system is correlated to the initial condition after an athermal, quasi-static (AQS) shearing protocol is applied. An experimental study on colloidal glasses by Chikkadi {\it et al.}~\cite{Chikkadi2014} reported bimodality for the spatial distribution of an order parameter constructed with the time-integrated mean-square displacement of particles. It is also worth to recall some other studies on glasses under shear~\cite{LamaitreCaroli,TanguyLanforteBarrat,Heussinger10}, in which a nontrivial statistics has been observed in the nonaffine displacements of particles, whose probability distribution exhibits peaks in different displacement ranges dependently on the observation time.
The present investigation differs from previous studies in an important way: The results displayed in Fig.~\ref{fig:bim} and Fig.~\ref{fig:usBim} show the succession in \emph{time} of two metastable states at $\sigma_{\mbox{\tiny{p}}}/\sigma_{\mbox{\tiny{Y}}} \simeq 1.4$ corresponding to different rheological branches. In other words, the whole system spends roughly the same amount of time in both the elastic and fluidized phases, constantly tunneling back and forth from one state to the other. Transitions are due to plastic events which eventually drive the system from the solid to the fluidized branch. Once the system reaches the fluidized branch, it flows plastically with a large number of plastic rearrangements (see Fig.~\ref{fig:deltaRegimes}). Plastic flow dissipates energy quite efficiently, and eventually, the power input due to the forcing is not able to sustain the energy dissipation due to plastic flow and the system goes back to the solid branch. Last but not least, we argue that the choice of heterogeneous stress enhances the probability to perform transitions between the two branches because this choice reduces the region (in physical space) where the system may switch form a flowing regime to a solid/elastic state (and vice versa).
This phenomenology differs from the bimodality discussed by Chikkadi {\it et al.}~\cite{Chikkadi2014}, since that is related to bimodality ``in space'' of the underlying shear. Our transitions in time, between elastic and fluidized states, also differ qualitatively from the observations of Refs.~\cite{Jaiswal2016,Procaccia17} on the intermittent periods of elastic loadings displayed in the failure of amorphous solids. Indeed, the loading and the failure take place on remarkably different timescales, which leads to a power-law distribution of the displacement field rather than a bimodal distribution (see Ref.~\cite{Benzi16} for a study of our model under shear flow). It must be also emphasized that in Ref.~\cite{Jaiswal2016} bimodality is reported for the overlap variable describing how well the system \emph{remembers} its initial configuration as a function of the applied quasistatic deformation: Such a choice would not allow to probe whether or not a given system repeatedly tunnels from a jammed to a flowing state and back, since the overlap is measured with respect to the starting configuration; thus, attaining a high overlap after a low value is reached is highly improbable. On the other hand, we observe bimodality for the time evolution of a rheological observable, signaling repeated transitions in time from a jammed to a flowing state and back, both states being unimodal in space.
The presence of bimodality in time, for both $\log[\Delta_S(t)]$ and $\log[{\hat u}_s(t)]$, should be related to long-range space correlations of plastic events, of the order of the domain size. In fact, for systems with a short-range space correlation, the effect of a single plastic rearrangement is unable to develop a cascade (in space and time) of other plastic events and trigger the transition of the {\it whole} system from the metastable solid branch to the metastable fluidized branch. A similar reasoning applies for the reversed transition: Once plastic rearrangements stop occurring in some part of the system, the flow ceases locally and the transition to the solid branch for the \emph{whole} system necessitates a correlation length that allows to cover the entire system size. This picture is actually borne out by a direct calculation of the correlation. A simple and intuitive way to look at space correlations is to compute the overlap-overlap correlation $G(r)$ that was already used in Ref.~\cite{ourSM14}: We follow the analysis presented in Ref.~\cite{Cavagna}, based on the idea of Ref.~\cite{LancasterParisi}. The physical meaning of $G(r)$ is rather clear. In a nutshell we can say that small values of $G(r)$ indicate that a part of the system moves somewhere while some other parts do not; large values of $G(r)$ mean the opposite, implying that different parts of the system move or not move at the same time. In other words, for large values of $\int \mbox{d}r\,G(r)$ (also known as dynamic heterogeneity~\cite{BerthierBiroli01}) the system either moves everywhere or does not move almost everywhere. We compute $G(r)$ as follows: we consider two times $t$ and $t+T_q$ and at each time we define the field $\phi(\vec{x},t) = \rho_A(\vec{x},t)-\rho_B(\vec{x},t) - \langle \rho_A-\rho_B \rangle_{\vec{x}}$, where $\langle \ldots \rangle_{\vec{x}}$ stands for space average and $\rho_A$, $\rho_B$ are the densities of the continuous and dispersed phases. Then, we define the overlap $q(x,y,t,t+T_q)$ as:
\begin{equation}
q(\vec{x},t,t+T_q) = \frac{\phi(\vec{x},t) \phi(\vec{x},t+T_q)}{  [ {\langle \phi(t)^2 \rangle_{\vec{x}}} {\langle \phi(t+T_q)^2 \rangle_{\vec{x}}} ]^{1/2} }. \label{overlap}
\end{equation}
Using Eq.~\eqref{overlap} we define the overlap-overlap correlation function, centered in the middle of the channel at $y=L/2$:
\begin{equation}
G(r) = \langle q(x,L/2,t,t+T_q) q(x,L/2+r,t,t+T_q) \rangle_{t,x} \label{corrover}
\end{equation}
where $\langle \ldots \rangle_{t,x}$ stands for time and $x$ averages and $T_q$ is chosen to be of the order of the time needed to perform a plastic rearrangement~\cite{ourSM14}. In Fig.~\ref{fig:corr} we show $G_c(r)$ (the connected part of $G(r)$) for $\sigma_{\mbox{\tiny{p}}} /\sigma_{\mbox{\tiny{Y}}} \simeq 1.2, 1.4$ and $1.6$. Clearly, at the transition point $\sigma_{\mbox{\tiny{p}}}/\sigma_{\mbox{\tiny{Y}}} \simeq 1.4$, $G_c(r)$ is very large everywhere in the system. It is crucial to remark that the correlation length observed for $\sigma_{\mbox{\tiny{p}}}/\sigma_{\mbox{\tiny{Y}}} \simeq 1.4$ differs from the cooperative scale $\xi$ of the system, the latter being equal to few droplet diameters~\cite{ourSM14}. When the system is in the fluidized branch for $\sigma_{\mbox{\tiny{p}}}/\sigma_{\mbox{\tiny{Y}}} \simeq 1.6$, the function $G(r)$ decays to zero with a correlation scale of the order of $\xi$~\cite{Goyon08,Goyon10}. These features in our model have already been observed in conditions of imposed shear in Ref.~\cite{ourSM14}. Here they are confirmed in a setup with imposed heterogeneous stress. Moreover, stress-controlled experiments (like the one we propose) somehow offer valid alternatives to the shear controlled ones~\cite{Varnik03,Chaudhuri12} in order to investigate the presence of multiple rheological branches. Indeed, if the system shows long-range correlations~\cite{KEP09} among plastic events, it may well be that in a shear-controlled experiment, the shear bands (if they form) are strongly fluctuating both in time and space. Eventually, these fluctuations would simply disappear in the average flow profile and one should rather observe a complex dynamics in time of the shear stress characterized by a strong intermittency of the time derivative of the stress, a phenomenology well reminiscent of the stick-slip behavior~\cite{Varnik03,Pignon96,Picard02}. In favor of this argument, we can mention the study by Varnik {\it et al.} on a model glass~\cite{Varnik03}, where the authors find that long-lived shear bands are replaced by the emergence of stick-slip phenomena with intermittent bursts; this happens at very low shear rates, i.e., at the point of discontinuity between the solid branch at $S=0$ and the fluid branch. We also mention the study by Pignon {\it et al.}~\cite{Pignon96} and by Picard {\it et al.}~\cite{Picard02} where the two regimes of stick-slip and shear bands are observed for different apparent shear rates; however, one has to notice that the stick-slip observed here is rather an oscillatory flow with undetectable intermittency. In the specific case of the theoretical model by Picard {\it et al.}~\cite{Picard02} this may be possibly related to the minimalistic nature of the model; i.e., no noise is added~\cite{ourSM16}.

In order to stress the combined role of multiple rheological branches and space correlations, it is worthwhile to further connect our observations with some other results presented in the literature~\cite{Chaudhuri_prl12,ChaudhuriHorbach14}. A recent work by Chaudhuri {\it et al.}~\cite{Chaudhuri_prl12} studied the interplay between the system size and the cooperative length in the flow arrest. Specifically, the model is that of soft-jammed repulsive disks (the ``Durian'' model~\cite{Durian}) in a periodic flow setup with heterogeneous stress, very similar to our stress profile. Upon decreasing the driving force, the authors determine the yielding threshold at which the flow ceases: Interestingly, under the conditions of periodic flow~\cite{Chaudhuri_prl12}, when the cooperative length becomes of the order of the system size, the authors find that the yielding threshold is increased with respect to the yield stress $\sigma_{\mbox{\tiny{Y}}}$, somehow in line with our findings (see Fig.~\ref{fig:rheology}). However, although an increased intermittency is reported at the onset of flow, the authors in Ref.~\cite{Chaudhuri_prl12} do not report any signature of metastable states like the one we observe, whereas simulation results are well predicted by the stationary fluidity model~\cite{KEP09}. This contrasts with our observations. The interplay between system size and cooperative scale was also highlighted in another work by Chaudhuri \& Horbach \cite{ChaudhuriHorbach14}, studying the transition to the flowing regime in a pressure-driven flow for a Yukawa binary fluid \cite{Zausch08,Zausch09}. When the cooperative length is of the order of the system size, it is shown that (in the long time limit) the system fluidizes nearly homogeneously. This behavior bears similarities with the transition from the solid-to-fluidized branch that we observe (see Fig. \ref{fig:three_Us}), with an important difference: The study by Chaudhuri \& Horbach \cite{ChaudhuriHorbach14} does not report the existence of metastable states; i.e., once the fluidized state is reached it is shown to persist for the whole simulation time. However, the time spent by the system in the {\it solid} phase is remarkably long, much longer than the time that would be observed for an \emph{unstable} state. In other words, one may argue that in Ref.~\cite{ChaudhuriHorbach14} two metastable branches coexist although the possibility of transition between the branches has not been investigated in detail. According to the results shown in the previous section and to the overlap-overlap correlation function shown in Fig. \ref{fig:corr}, we identify two conditions that should be satisfied for a clear signature of metastable states: there must exist a difference between the static and the dynamic yield-stress values (i.e., there must exist two rheological branches) and there must be long-range correlations among plastic events. In Refs.~\cite{Chaudhuri_prl12,ChaudhuriHorbach14} it is unknown whether one or both requirements are not met. We may argue that the model used by the authors in Ref.~\cite{Chaudhuri_prl12} is rather a model for a nonadhesive emulsion~\cite{Mansard13}, and the difference between the static and dynamic yield stress is so small~\cite{Chaudhuri12} that metastability between two different rheological branches cannot be observed. The Durian~\cite{Durian} model has also been used recently by Kawasaki \& Berthier~\cite{KawasakiBerthier2016} to study the yielding transition under oscillatory flow. By analyzing the displacement fields of the particles the authors report a rather discontinuous transition at the yield stress: While above yield the fluctuations in the displacement fields are persistent in time (fluidized state), below the yield stress they are metastable and cease after some time. The possibility of transitions back to the fluidized state has not been studied in detail when changing the stress protocol and/or for longer simulation times, but again we argue that it would not be observed because of the model used~\cite{Chaudhuri_prl12}. All these considerations suggest that studies regarding the presence of shear bands and ``stick-slip'' should be consistently accompanied with measurement of the correlation functions. Correlations of the microscopic strain field were actually measured by Chikkaddi {\it et al.}~\cite{Chikkadi2011} in colloidal glasses showing the formation of shear bands; however, such results were only obtained for the two bands separately.

\begin{figure}[b]
\includegraphics[scale=0.76]{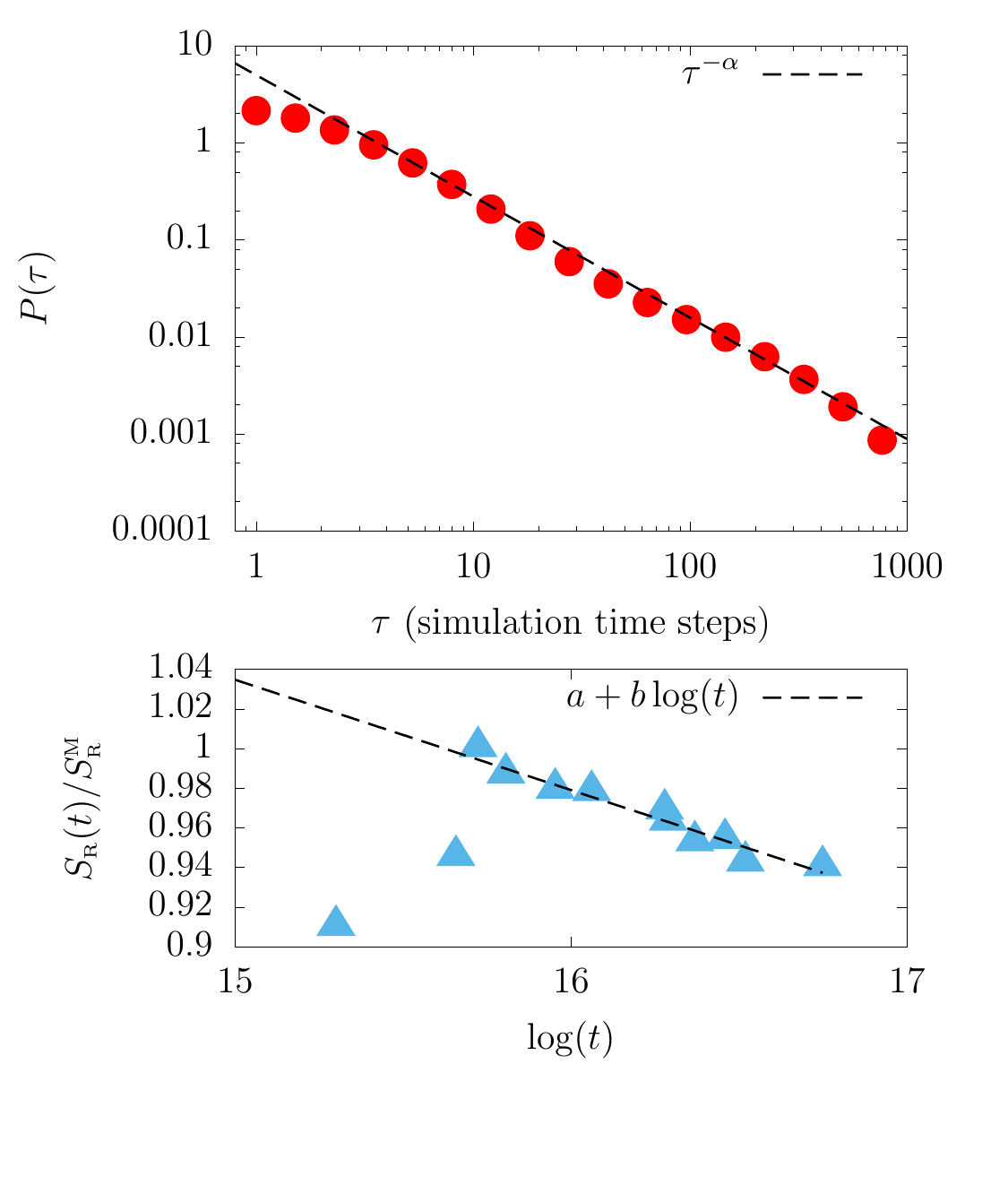}
\caption{\label{fig:attesa} Top panel: Probability distribution for the trapping time
  $\tau$ (in simulation time steps), i.e., the time spent in the solid state, at $\sigma_{\mbox{\tiny{p}}}/\sigma_{\mbox{\tiny{Y}}} \simeq 1.4$. The distribution can be well fit by a power law $P(\tau)\sim \tau^{-\alpha}$ with $\alpha\sim 1$ indicating a self-similar structure in the intermittent transitions from the solid to the fluidized state. Such self-similar distribution has also been measured in spin glasses~\cite{Berthier2003}. In the lower panel we show the running average $S_{\mbox{\tiny{R}}}(t)$ of the shear $S(t) = 2\pi u_s(t)/L$, normalized to its maximum $S_{\mbox{\tiny{R}}}^{\mbox{\tiny{M}}}$ (see text for details). The behavior of $S_{\mbox{\tiny{R}}}(t)$ is consistent with a logarithmic decay.}
\end{figure}

Further analysis in our numerical simulations is also stimulated by a direct comparison of the phenomenology that we observe to that of glassy models~\cite{Varnik03,Varnik04,Berthier2003} and, in particular, finite size $p$-spin models~\cite{Berthier2003}. The nontrivial and interesting point is the observation that the system spontaneously develops two stable branches in its phase-space dynamics, similarly to the two rheological branches needed to describe the formation of shear bands. Such systems are also known to display a dynamic transition at some temperature $T_d$. For $T<T_d$ the system is trapped in a large number of states, which grows as the exponential of its size. Upon applying an external force, the system shows a dynamic transition similar to a yield-stress transition. For a finite number of spins, the system exhibits bursts of activity, i.e., the activated process, which show self-similarity in size and time~\cite{Berthier2003,Cugliandolo97}. The probability distribution of the trapping time $\tau$, namely the time between two successive bursts, shows a scaling behavior $P(\tau) \sim \tau^{-a}$ with $a = 1+T/T_d$. This behavior is qualitatively similar to the one described by SGR theories~\cite{Sollich97,Sollich98,Sollich2000} based on the trap model~\cite{bouchaud1992weak}. Going back to our results, for the case where $P(\log(\Delta_{\mbox{s}}))$ is bimodal ($\sigma_{\mbox{\tiny{p}}} /\sigma_{\mbox{\tiny{Y}}} \simeq 1.4$), we can define the trapping time $\tau$ spent by the system in the solid branch: We use the value of $\log(\Delta_{\mbox{s}})$ at the local minimum (see Fig.~\ref{fig:bim}) as a threshold to condition the data. We expect $\tau$ to be a random variable and we look at the probability distribution $P(\tau)$ shown in Fig.~\ref{fig:attesa}. The probability distribution $P(\tau)$ behaves as a scaling function of $\tau$, i.e. $P(\tau) \sim \tau^{-\alpha}$ with $\alpha \sim 1$ showing the existence of nontrivial time correlations. In the bottom panel of Fig.~\ref{fig:attesa} we show the running average $S_{\mbox{\tiny{R}}}(t)$ of the shear $S(t) = 2\pi u_s(t)/L$ [see Eq.~\ref{eq:viscousProj} and below], normalized to its maximum $S_{\mbox{\tiny{R}}}^{\mbox{\tiny{M}}}$ for the bimodal forcing $\sigma_{\mbox{\tiny{p}}}/\sigma_{\mbox{\tiny{Y}}} \simeq 1.4$. The running average $S_{\mbox{\tiny{R}}}(t)$  is  computed when the system is in the flowing phase, i.e., when $u_s(t)$ belongs to the larger peak shown in the middle panel of Fig.~\ref{fig:usBim}; thus the value of the minimum of $\log|\hat{u}_s|$ is used as a cutoff. From the bottom panel of Fig.~\ref{fig:attesa} it is possible to see that the time evolution of $S_{\mbox{\tiny{R}}}(t)$ is consistent with a logarithmic decay. Indeed, this is a further characterization of our results that can be verified in non-homogeneous stress experiments such as the one that we outline in the following section.

\section{Pressure-driven flows}\label{sec:confined}
\begin{figure}[t]
\includegraphics[scale=0.8]{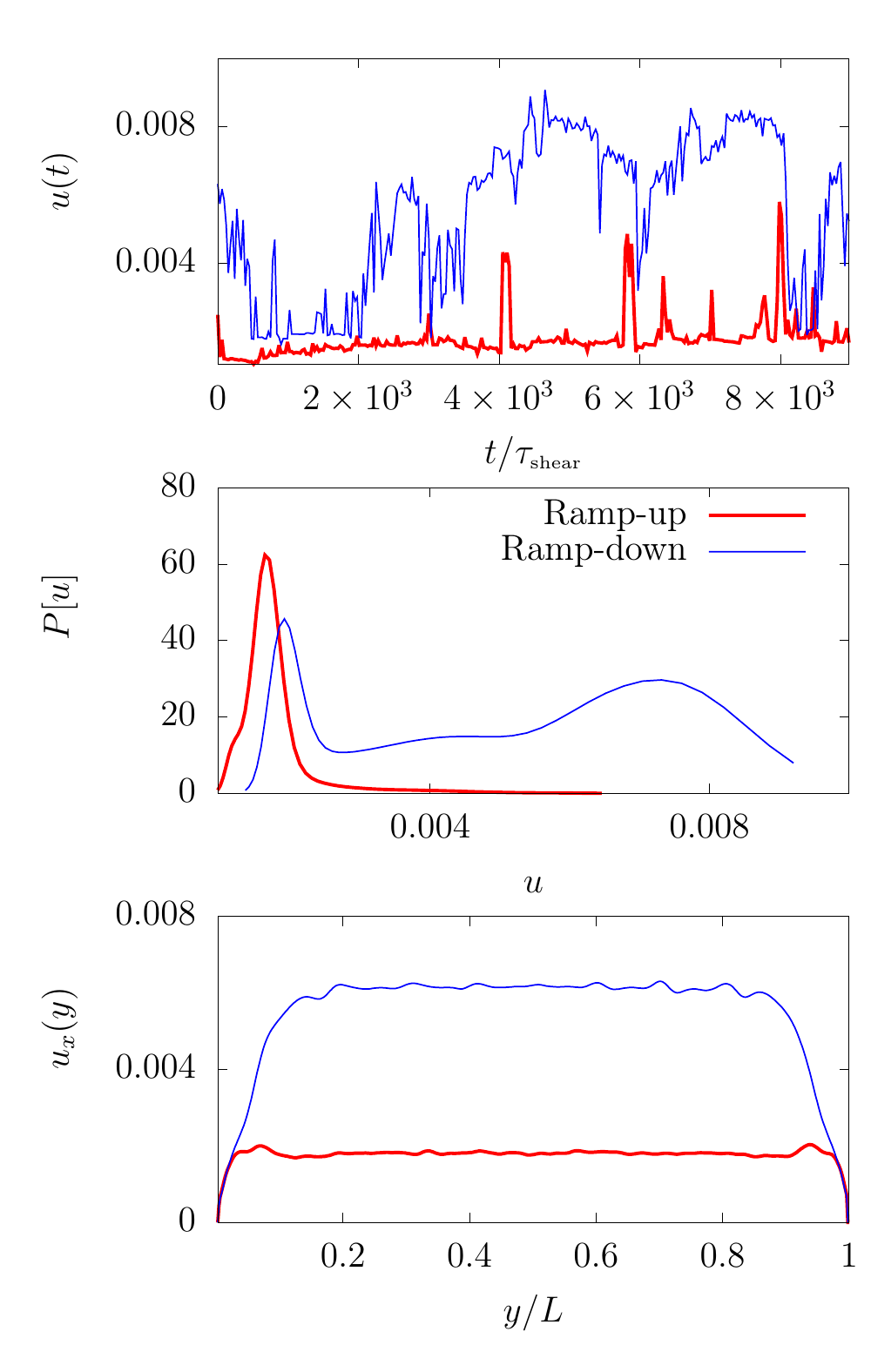}  
\caption{\label{fig:poisBim} Results for the flow driven by a constant pressure gradient producing a peak stress $\sigma_{\mbox{\tiny{p}}}/\sigma_{\mbox{\tiny{Y}}} \simeq 1.4$. Different line thicknesses (indicated also with different colors) correspond to different system preparations: Data reported using the thick red line refer to a system previously driven from a lower forcing (i.e., ramp-up protocol), whereas data represented by a blue thin line refer to a system previously driven at a larger forcing (i.e., ramp-down protocol). Top panel: Velocity flux $u(t)$ as a function of time normalized by the shear time $\tau_{\mbox{\tiny{shear}}}$. The thick red line displays some intermittent spikes while the blue one shows a sequence of transitions. Middle panel: Probability distributions for the velocity flux $u$ displaying a single peak for the thick red line and a bimodal character for the thin blue line. Bottom panel: Velocity profile for $u_x(y)$ averaged over time and along the streamwise direction $x$, the thick red curve indicates a plug-flow dynamics dominated by an elastic bulk, while the thin blue one shows a developed velocity gradient near the boundaries.}
\end{figure}

The results discussed in the previous sections refer to fully periodic boundary conditions. In this section we want to comment about the possibility to obtain the same results in the case of realistic boundary conditions. In particular we consider the case of a pressure-driven flow in a two-dimensional channel and streamwise periodic boundary conditions. Since the system is driven with a constant force (pressure gradient) in the streamwise direction, the stress is a linear function of the coordinate $y$ (see Fig.~\ref{fig:sketch}) and its absolute value reaches the maximum $\sigma_{\mbox{\tiny{p}}}$ at the boundaries. The system shows a rather clear apparent slip \cite{MeekerBonnecaze04,Seth12,Bonn15} at the smooth boundaries and this goes together with a nonzero mean flow; hence, the analysis in terms of $\Delta_{\mbox{s}}$ is no longer suitable. Furthermore, because of strong localization of plastic events at the boundary, there is less energy available to switch from one rheological branch to the other for the whole system. This implies that the characteristic ``trapping'' time becomes much longer with respect to the one observed in periodic boundary conditions. To perform long-time numerical simulations, we choose a square system with side $L=512$ lattice points. In Fig.~\ref{fig:poisBim} we show the most interesting information obtained from our simulations. We choose $\sigma_{\mbox{\tiny{p}}}/\sigma_{\mbox{\tiny{Y}}} \simeq 1.4$ and we run simulations imposing a pressure gradient on a configuration picked from a lower forcing steady state (i.e., ramp-up protocol). The interesting variable to look at is the velocity flux $u(t)$ defined as the space average at time $t$ of the stream-wise velocity. In the upper panel of Fig.~\ref{fig:poisBim} we show $u(t)$ (thick red line) for about $9\times 10^3$ shear times. The system shows a nonzero average velocity (due to the slip at the boundaries) with superimposed bursts of larger values, similar to a stick-slip behavior. The probability distribution of $u$ is shown in the middle panel, while the average velocity profile is shown in the bottom panel. Next, we increase $\sigma_{\mbox{\tiny{p}}}$ so that the system reaches a fluidized state (not shown). Once the statistical properties in the fluidized state could be considered stationary, we reduced the pressure gradient (i.e., ramp-down protocol) and perform a new numerical simulation at the same value of the peak stress $\sigma_{\mbox{\tiny{p}}} \simeq 1.4$ already discussed. For this new simulation the results are reported with the thin blue line in Fig.~\ref{fig:poisBim}. It is quite clear that the system shows transitions in the rheological behavior, characterized by small and large values of $u$ (see the probability distribution). The qualitative picture is similar to the one discussed in the previous section although the time scale is much longer.

The results shown in Fig.~\ref{fig:poisBim} can be considered a preliminary investigation for systems with realistic boundary conditions. The point we want to highlight here is that the existence of two metastable states, discussed in the previous section, can be observed numerically and (most importantly) experimentally with long-time statistics (order $10^4$ shear times of the system) and with a fine scanning of the forcing parameters. Moreover, further analysis is required to investigate hysteresis effects.

\section{Concluding Remarks}\label{sec:conclusions}
Based on numerical simulations of a soft-glassy model we have studied its rheological response with an imposed space-dependent stress in an ideal fully periodic setup. The rheological properties of the model under study show the existence of multiple rheological branches with a difference between static and dynamic yield stress. The peak value $\sigma_{\mbox{\tiny{p}}}$ of the imposed stress is set close to the static yield stress of the material. We observe that the time dynamics of the system is remarkably non-steady, as it tunnels intermittently between two different states, a ``solid'' state and a ``fluidized'' one. Numerical simulations~\cite{ourEPL10,ourSM12,ourEPL13,ourSM14,ourJFM15,ourEPL16} allow to bridge the rheological response at large scales to the behavior displayed deeper down at small scales, where we observe a bimodal probability distribution of the largest value of the displacement field describing recurrent transitions in time between two unimodal states in space. Our results highlight the role of plastic rearrangements as the mechanical trigger for the hopping between the two states as well as the role of long-range correlations for the hopping to occur. Preliminary investigations have shown that such scenario holds for the more realistic case of a flow driven by a constant pressure gradient. Hence, such nonsteady yielding dynamics with recurrent transitions can be put to test by laboratory experiments.

From a general perspective, we point out that the existence of multiple rheological branches has been often introduced to explain the formation of permanent shear bands in soft-glasses~\cite{Varnik03,Berthier2003}. From this point of view, the formation of shear bands can be considered as a ``phase separation'' in space, allowing the space coexistence of solid and fluidized regions~\cite{KEP09}. Our observations somehow take a broader perspective and generalize the idea of coexistence in the time domain. For this coexistence in time, both long-range space correlations and the stress protocol are crucial. We indeed argue that a spatial correlation length of the order of the system size is crucial to trigger transitions between states and establish the time coexistence. Moreover, we argue that the choice of heterogeneous stress enhances the probability to perform transitions between the two branches because this choice reduces the region (in physical space) where the system may switch form a flowing regime to a solid/elastic state (and vice versa).

Given the role of space correlations in our system, it is then natural to comment on their expected role in a ``classical'' shear-banding scenario, i.e., when heterogeneous flow is observed in presence of a shear-controlled experiments with homogeneous stress~\cite{BonnReviewYIELDSTRESS}. We indeed argue that the ``phase coexistence'' in space can be observed only if short-ranged correlations are present, whereas in presence of long-ranged correlations one would rather expect a stick-slip behavior. Noteworthy, preliminary simulations of our system under the conditions of imposed shear flow do not show permanent shear bands \cite{Benzi16}. The above discussion may suggest that some complex dynamic and rheological properties observed in some soft-glasses, namely stick-slip behavior~\cite{Pignon96,Picard02,Ianni08,Divoux11} and formation of permanent shear bands~\cite{Ovarlez09,Fielding2}, can somehow be unified within the same theoretical framework, dependently on the range of space correlations. Given this view, it could be interesting to revisit our recent proposal~\cite{ourSM16} where cooperativity effects have been linked to the formation of permanent shear bands. One could add to the model a tunable correlation between plastic rearrangements and explore the consequences on the formation of the bands.

We remark that our findings share many features with the analysis performed on $p$-spin glasses near the dynamic transition at the temperature $T=T_d$. The analysis performed in Ref.~\cite{Berthier2003} shows that for $T<T_d$ the system develops two stable rheological branches. Moreover, the trapping time in the solid branch shows a power-law distribution which is also observed in our system. Finally, the theoretical analysis in Ref.~\cite{parisiFranz2000} shows that, near the critical temperature, the system displays bimodality in the order parameter and long-range correlations in space (i.e., diverging dynamic heterogeneity), because of the spinodal character of the transition. All the above features are observed in our simulations.

\begin{acknowledgements}
  The research leading to these results has received funding from the projects ``High performance data network: Convergenza di metodologie e integrazione di infrastrutture per il calcolo High Performance (HPC) e High Throughput (HTC)'' (fondi CIPE) and MIUR-PRIN Grant No. 2015K7KK8L. Massimo Bernaschi is gratefully acknowledged for computational support.
\end{acknowledgements}

\bibliography{refsmod}
\bibliographystyle{apsrev4-1mod.bst}

\end{document}